\documentclass{article}    

\usepackage{graphicx}

\title{\textbf{Trans-Coordinate Physics}}  
\author{Richard Mould\footnote{Department of Physics and Astronomy, State University of New York, Stony Brook,
\mbox{New York} 11794-3800; http://ms.cc.sunysb.edu/\~{}rmould}}  
\date{}    
   
\begin{document}             

\maketitle              

\begin{abstract}

 Standard practice attempts to remove coordinate influence in physics through the use of invariant equations.  Trans-coordinate physics proceeds differently by not introducing space-time coordinates in the first place.  Differentials taken from a novel limiting process are defined for a particleÕs wave function, allowing the particleÕs dynamic principle to operate `locally' without the use of coordinates.  These differentials replace the covariant differentials of Riemannian geometry.  With coordinates out of the way `continuous  conservation principles' and the `Einstein field equation' are no longer fundamentally defined; although they are constructible along with coordinate systems so they continue to be analytically useful. Gravity waves as presently understood are not defined, so we conclude that the gravitational detectors LIGO and Weber bar will not work.  We assume that gravitons alone are fundamentally responsible for gravity.   It is shown how gravitational uncertainty can be reconciled with the certainty of quantum mechanical dynamics.

\end{abstract}

\section*{General features}
James Clerk Maxwell was the first to use space-time coordinate systems in the way they are used in contemporary physics.  They play a role in his formulation of electromagnetic field theory that makes them virtually indispensable.  Einstein embraced Maxwell's methodology but devoted himself to eliminating the influence of coordinates because they have nothing to do with physics.  However, his use of relativistic invariance for that purpose does not really eliminate the influence of coordinates, as will be evident below where they are removed \emph{entirely} from physics.  

Trans-coordinate physics proceeds on the assumption that space-time coordinates should not be introduced at any level.  As a practical matter, and for many analytic reasons, coordinates are very useful and probably always will be.  But if nature does not use numerical labeling for event identification and/or analytic convenience, and if we are interested in the most fundamental way of thinking about nature, then we should avoid space-time coordinates from the beginning. 

Without coordinates, relativity physics resides solely in the properties of the embedding metric space, and quantum mechanics resides solely in properties of local wave functions that are assigned to particles.  These two domains overlap `locally' where a Lorentz invariant quantum mechanics is assumed -- a more general invariance for the dynamic principle is not required in this physics.  Massless particles that move in the space `between' massive particles have a reduced definition and function.

As a result, the variables of a particle's wave function are wholly contained inside the wave packet and are coordinate independent.  They move with a particle's wave function in the embedding metric space, but they do not locate it in that space.  No particle has a \emph{net velocity} or \emph{kinetic energy} when considered in isolation, for these quantities require a coordinate framework for their definition.  This alone reveals the radical nature of removing coordinates \emph{entirely} from physics.  

Another consequence of this program is that energy and momentum are not propagated through the empty space between particles.  Although particle energy, momentum, and angular momentum are conserved in local interactions, we say that nature does not provide for conservation in the space between them.  We are the ones who make these provisions through our introduction of regional coordinates that we use to give ourselves the big picture.  Doing so facilitates analysis.  

Before applying these ideas to a fully quantum mechanical and relativistic system, we will look at a Newtonian inertial system with and without coordinates  in order to get an understanding of the relationship that exists between coordinates and the dynamic conservation principles.

\section*{A Newtonian inertial system}
ANewtonian inertial system contains a number of space-time symmetries that exist prior to the introduction of coordinates.  First, a clock in that space ticks the same no matter what its position, or state of motion, or time.  It is forever the same. Second, a meter stick will have the same length no matter when or where it is used, independent of its orientation or state of motion.  Against these `background' symmetries, all of the objects in the system are subject of Newton's laws; so in any collision between particles, their energy, momentum, and angular momentum is conserved.  

If no coordinates are allowed, then there is no way that velocity can be defined for a single isolated particle, and this means that a  particle cannot carry energy, momentum, or angular momentum away from (or toward) a collision.  Granting that it can conserve these quantities during an interaction, a particle cannot have these properties between collisions without the benefit of something completely artificial like a coordinate system or, in Newton's case, an absolute space.  

The above symmetries allow us to construct Cartesian  coordinates  together with a single universal time.  Each particle can then be given an energy, momentum, and angular momentum \emph{at every moment of time}, not just during collisions.  The Noether theorem, requiring that ``each space-time symmetry gives rise to a conservation principle'', is satisfied.  So the temporal symmetry involving clocks implies the conservation of energy, the displacement symmetry involving meter sticks implies the conservation of momentum, and the rotational symmetry implies the conservation of angular momentum.

It is the intent of trans-coordinate physics to do the same for a relativistic quantum mechanical system.  In the following we will remove coordinates from the formulation of fundamental processes.  There will be many similarities between the result of this removal and those described in the above Newtonian system.  We will find for instance, that the conservation principles hold during particle interactions in these systems, but that energy, momentum and angular momentum (apart from spin) are \emph{not defined} between collisions.  Also, as in the Newtonian case, space-time symmetries exist is special situations, so it is often possible to define coordinate systems in which these conserved quantities are continuously conserved between interactions.  Hence the great usefulness of coordinates!  

The thesis of this chapter is that \emph{nature does not need or use coordinates}.  We are the ones who introduce these props for our own purposes; and as a result, most of field theory must be fundamentally abandoned.  This leads to a reformulation of electromagnetic theory and results in the demise of general relativity. It is concluded that general relativity is a macroscopic approximation to a more correct graviton theory, thereby resolving the clash between the otherwise mismatched disciplines of general relativity and quantum mechanics \cite{tH, JM}. 
The treatment in this paper is primarily concerned with electromagnetic theory and relativity.  We begin by assuming the existence of the invariant metric background of general relativity -- the canvas of creation.

\section*{Partition lines}
The first thing we will do is establish the local validity of the dynamic principle in quantum mechanics.  Since the dynamic principle is a differential equation, careful attention must be given to how these derivatives are defined without using coordinates.

In Minkowski space one must choose a world line in order to define a time coordinate of an event \textbf{a}.  For a massive particle  it will be shown  possible to choose a unique world line at each location inside the particle's wave packet that is specific to the particle at that location.  That world line corresponds to the direction of \emph{square modular flow} at that event.  The collection of these world lines over the particle's wave packet can be thought of as the streamlines of its square modular flow in space and time.  They are called \emph{partition lines}.  We will also define \emph{perpendiculars} that are space-like lines drawn through each event perpendicular to the local partition line.  We will first develop the properties of partition lines in a 1 + 1 space, and then in 2 + 1 and 3 + 1 spaces.

\begin{figure}[b]
\centering
\includegraphics[scale=0.8]{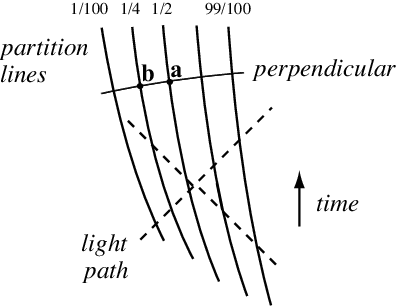}
\center{Figure 1: Partition lines in a Minkowski space}
\end{figure}

Figure 1 is a 1 + 1 Mnkowski surface with light paths given by $45^\circ$ dashed lines.  Partition lines of an imagined particle wave packet are represented in the figure by the five slightly curved and more-or-less vertical lines.  They tell us that the wave packet moves to the left with ever decreasing velocity and that it spreads out as it goes.  This description is not trans-coordinate because it is specific to the Lorentz frame in the diagram; but these lines provide a scaffold on which it is possible to hang a trans-coordinate wave function. 

Partition lines pass through every part of the particle's wave packet and do not cross one another.  They are not defined outside of a wave packet.  Just as the space is initially given to us in the form of a metric background, any particle is initially given in the form of partition lines with the above characteristics.  A more exact interpretation of these lines is given below where values are assigned to them in a way that reflects the intended \emph{given conditions}.  These conditions are not `initial' in the usual temporal sense, but are `given' over the space-time region of interest.

Let the third partition line from the left (i.e., the middle line in Fig.\ 1) portion off 1/2 of the packet, so half of the particle lies to the left of an event like event \textbf{a} in the figure.  That is, there is a 0.5 probability that the particle will be found on the perpendicular (defined in the next section) extending to the left of \textbf{a}.  This statement is assumed to have objective invariant meaning.  Of course, the other half of the particle lies to the right of event \textbf{a} on the perpendicular through \textbf{a}.  The middle partition line is made up of all the events in the wave packet that satisfy this condition, so they together constitute a continuous line to which we assign the value of 1/2.  There is a 0.5 probability that the particle will be found \emph{somewhere} on the left side of this line. 

In a similar way we suppose that the second partition line in Fig.\ 1 portions off, say, 1/4 of the packet on the perpendicular to the left of an event \textbf{b}, and that the first line portions off 1/100 of the particle or some other diminished amount.  We further assume that the fifth line goes out to 99/100 of the particle packet, so the entire particle is represented by streamlines that split the particle into objectively defined fractional parts. 

When a wave function is finally assigned we will show that its total square modulus remains `constant in time' between any two partition lines in  1 + 1 space, and is similarly confined in higher dimensions.

\section*{Neighborhoods}
Every event inside the wave packet has a unique time direction defined for it by the partition line passing through the event.  This allows us to define unique \emph{inertial} neighborhoods associated with each event.

Consider a flat space inside the wave packet of a massive particle and the Minkowski metric that is intrinsic to that space.  Beginning with an \mbox{event \textbf{a}} in Fig.\ 2a, proceed up the particle's partition line through \textbf{a} by an amount \mbox{-$\Delta$} which is the magnitude of the invariant interval from event \textbf{a} to an event \textbf{b}.  This interval \textbf{ab} is negative and identifies the chosen time axis inside the particle packet at event \textbf{a}.  Then find event $\textbf{b}'$ by proceeding down the partition line the same invariant interval -$\Delta$.  Construct a backward time cone with \textbf{b} at its vertex and a forward time cone with $\textbf{b}'$ at its vertex and identify the intersection events \textbf{c} and $\textbf{c}'$.  Since these events are embedded in a one dimensional flat space, the positive space-like interval $\textbf{cc}'$ will pass through event \textbf{a} and will be bisected by it with

\begin{displaymath}
\textbf{ca} = \textbf{ac}' = \textbf{cc}'/2 = \Delta > 0
\end{displaymath}
For any $\Delta$, all of the events included in the intersection of the light cones of \textbf{b} and $\textbf{b}'$ are defined to be a \emph{neighborhood} of event \textbf{a}.  The events along the \emph{perpendicular line} $\textbf{cc}'$ are defined to be a \emph{spatial neighborhood} of \textbf{a}.  The limit as $\Delta$ goes to zero is identical with the limit of small neighborhoods around \textbf{a}.

\begin{figure}[t]
\centering
\includegraphics[scale=0.8]{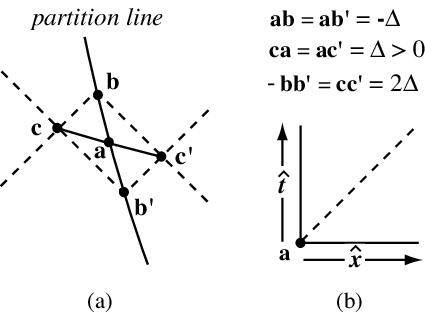}
\center{Figure 2: Establishing neighborhoods}
\end{figure}

\section*{Curved space}

The above considerations for a `flat' space also apply \emph{locally} in any curved space, so we let the conditions in Fig.\ 2a be generally valid in the limit as $\Delta \rightarrow 0$.  Figure 2b shows the resulting Minkowski diagram in the local inertial system with $\hat{x}$ and $\hat{t}$ as the space and time unit vectors in the directions $\textbf{ac}'$ and \textbf{ab} respectively. 

The unit of these vector directions is that of $\surd\Delta$ in meters, although we have not established coordinates along those directions.  Specifically, we have not established a unique numerical value attached to an event \textbf{a} or a distant zero-point for that value; so the development so far is consistent with the trans-coordinate (or coordinateless) aims of this paper.

Unit vectors at event \textbf{a} will be referred to as the \emph{local grid} at event \textbf{a}, where the time direction is always along the partition line going through \textbf{a}, and the spatial direction is clockwise along the perpendicular.  These definitions have nothing to do with the curvature of the space in the wave packet at or beyond the immediate vicinity of \textbf{a}.  Every event inside a particle packet has a similar local grid.  The local grids of other events in the neighborhood of event \textbf{a} will be continuous with the local grid at a in this 1 + 1 space, but not for higher dimensions as we will see.

\section*{The wave function}
We specify the quantum mechanical wave function at each event \textbf{a} in a particle wave packet over the space-time region of interest
\begin{equation}
\psi(\textbf{a})
\end{equation}
which is identified in the manner of Euclid's geometry since there are no coordinate numbers involved.   There are four auxiliary conditions on this function.

\vspace{.1cm}
\noindent
\textbf{First}: The function $\psi(\textbf{a})$ is a complex number given at event \textbf{a} that is continuous with all of its
neighbors.  The unit of $\psi$ is $m^{-1/2}$ in this 1 + 1 space.

\noindent
\textbf{Second}: Partial derivatives of $\psi(\textbf{a})$ are defined in the  limit of small neighborhoods around \textbf{a} (i.e., for small values of  $\Delta$).  
\begin{eqnarray}
\partial\psi(\textbf{a})/\partial x &=& \lim_{\Delta\rightarrow 0}  \frac{\psi(\textbf{c}') - \psi(\textbf{c})}{2\sqrt{\Delta}} 
\\
\partial\psi(\textbf{a})/\partial t &=& \lim_{\Delta\rightarrow 0}  \frac{\psi(\textbf{b}) - \psi(\textbf{b}')}{2\sqrt{\Delta}}
\nonumber 
\end{eqnarray}
The second spatial derivative  is then
\begin{displaymath}
\partial^2\psi(\textbf{a})/\partial x^2 = \lim_{\Delta\rightarrow 0}  \frac{\partial\psi(\textbf{c}')/\partial x -
\partial\psi(\textbf{c})/\partial x}{2\sqrt{\Delta}} 
\end{displaymath}

Notice that we have defined derivatives in the directions $\hat{x}$ and $\hat{t}$ without using coordinates to `locate' or numerically `identify' events along either of those directions.  Only $\Delta$-\emph{intervals} are taken in the limit from the invariant metric space.

\vspace{.1cm}
\noindent
\textbf{Third}: The value of $\psi$ at event \textbf{a} is related to its neighbors  through the
\emph{dynamic principle}.  This principle determines how    $\psi(\textbf{a})$ evolves relative to its own time against the metric background, and how it relates spatially to its immediate neighbors.

\vspace{.1cm}
\noindent
\textbf{Fourth}: The objective fraction of the particle found between the partition line through event \textbf{c} in Fig.\ 2a and a
partition line through event $\textbf{c}'$ is equal to $f_{cc'}$.  In the limit as $\textbf{cc}'$ = $ 2\Delta$ goes to zero, the fraction
of the particle between  differentially close partition lines goes to zero.  Normalization of $\psi(\textbf{a})$ is stictly local and requires  
\begin{displaymath}
\psi^*(\textbf{a})\psi(\textbf{a}) = \lim_{ \Delta \rightarrow\ 0}\frac{f_{cc'}}{2\Delta}
\end{displaymath}

\noindent
It follows that
\begin{displaymath}
 \psi^*(\textbf{a})\psi(\textbf{a}) = \psi^*(\textbf{b})\psi(\textbf{b}) = \psi^*(\textbf{b}')\psi(\textbf{b}') 
\end{displaymath}
because the fractional difference between any two partition lines is the same over any perpendicular.  Therefore, the square modular flow will be \emph{constant in time} between any two partition lines as previously claimed.

These four auxiliary conditions must be satisfied when taken together with the initially given partition lines, but there is no guarantee that there exists a wave function that qualifies.  \emph{Finding a solution} therefore consists of varying the partition lines (i.e., the given conditions) until a wave function exists that satisfies these conditions.

The choice of a world line based on partition lines is not a coordinate choice, nor is the accompanying limiting procedure.  So these definitions are not just coordinate invariant, they are fully \emph{coordinate free}.  They allow us to find physically creditable derivatives of any continuous function in a way that is entirely local and independent of curvature, and to found physics on that basis.

\section*{One particle}
Partition lines do not extend beyond the particle, so in the absence of `external' coordinates that extend beyond the particle (in an otherwise empty space) there is no basis for claiming that the particle has a net velocity, kinetic energy, or net momentum.  This will be true of both zero and non-zero mass particles.  It is a consequence of a trans-coordinate physics that particles only manifest energy and momentum when interacting with other particles.

It is not meaningful to say that the function $\psi(\textbf{a})$ at a single event inside a particle is a superposition of eigenstates.  That is coordinate language that may be analytically useful but is not basically applicable.  Trans-coordinate language concerning $\psi(\textbf{a})$ is strictly local, saying that each $\psi(\textbf{a})$ and its derivatives satisfy the dynamic principle \emph{at that event}.   Similarly, each part of the particle's wave packet follows its own world line, so the there is no single world line for the particle as a whole, as is apparent in Fig.\ 1.  Apart from the postulated background metric and the regional definition of the partition lines, nature does not drive the particle as a whole.  It deals separately with each part.  Several exceptions to this are described in the following  and are fully invariant.

\section*{Internal coordinates}
We want the wave function $\psi(\textbf{a})$ in Eq.\ 1 in a form that permits internal analysis.  So to give ourselves an internal picture of the particle we use the grid we established at each event to construct internal coordinates.  To do this starting at event \textbf{a}, integrate the negative square root of the metric along the partition line going through \textbf{a} and assign a time coordinate $t$ with an origin at \textbf{a}.  Then integrate the square root of the metric over the perpendicular going through event \textbf{a} and assign a space coordinate $x$ with an origin at \textbf{a}.  The coordinates $x$ and $t$ may be extended over the entire object yielding a wave function that can be written in the conventional way $\psi(x, t)$.  These internal coordinates are created by us for the purpose of analysis and understanding.  They have no natural significance.

With internal coordinates we can integrate across one of the perpendiculars to find the \emph{width} of the wave packet.  It should also be possible to integrate the square modulus over a perpendicular to find the \emph{total normalization}.  That total will be equal to 1.0 if $df$ is equal to the fraction of the particle sandwiched between two differentially close partition lines as claimed.  We can also use internal coordinates to define the internal variables of a particle, including energy and momentum eigenstates.  Quantities such as width and total normalization are our creation and have no fundamental status.

\section*{Three and four dimensions}

Imagine that a particle's wave packet occupies the two-dimensional area shown on the space-like surface in Fig.\ 3.  The surface is divided into a patchwork of squares, each of which is made to contain a given fraction of the particle like $1/100^{th}$ of the particle.  Each of these squares has four distinguishable crossing points or corners.  A similar two-dimensional patchwork is constructed on all of the space-like surfaces through which the particle passes in time, thereby creating a continuous 2 + 1 scaffold.  Each of the enclosed areas generated in this way is required to contain 1/100 of the particle, and its corners are penetrated by the partition lines of the particle.  As in the 1 + 1 case, these lines may be thought of as streamlines of the square modular flow of the particle through time.  In the limit as this fraction goes to zero, partition lines pass through each event on the space-like surface in the figure and they do not cross one another.

\begin{figure}[t]
\centering
\includegraphics[scale=0.8]{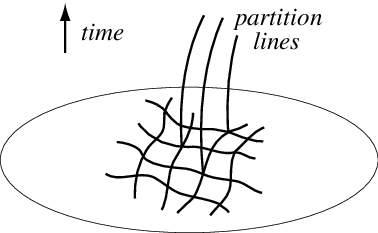}
\center{Figure 3: Two dimensional scaffold}
\end{figure}

Space-time directions are chosen for a given partition line in a way that is similar to the procedure in Fig.\ 2.  Starting with an event \textbf{a} in Fig.\ 4a, move up its partition line a metrical distance -$\Delta$ to event \textbf{b}.  Then find $\textbf{b}'$ by proceeding down the partition line the same invariant interval -$\Delta$.  Construct a backward time cone with \textbf{b} at its vertex and a forward time cone with $\textbf{b}'$ at its vertex and identify their intersection in the space-like two-dimensional loop shown in \mbox{Fig.\ 4a}, in the limit as $\Delta$ goes to zero.   In the local inertial system, two perpendicular unit vectors $\hat{x}$ and $\hat{y}$ are chosen along the radius of the circle of radius $\Delta$ that spans the spatial part of the grid at event \textbf{a}.  For any $\Delta$, choose a space-like line going through \textbf{a} that is aligned with $\hat{x}$ and extends to the circumference of the circle in Fig.\ 4a.  This line will have a single + intercept and a single - intercept, and does not have to be `straight' so long as it is aligned with the unit vector at event \textbf{a}.  The same will be true of $\hat{y}$ intercepts.

\begin{figure}[b]
\centering
\includegraphics[scale=0.8]{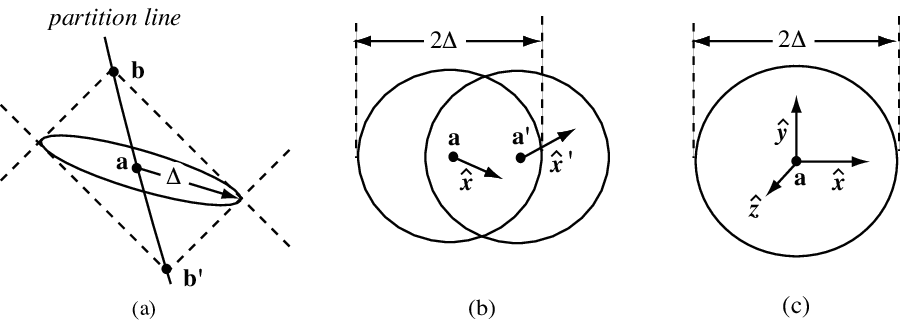}
\center{Figure 4: Establishing space-like unit vectors}
\end{figure}

The spatial grids of nearby events such as \textbf{a} and $\textbf{a}'$ in Fig.\ 4b do not have to line up in any particular way.  Even if they are in each other's spatial neighborhood for some value of $\Delta$, their unit vectors $\hat{x}$ and $\hat{x}'$ will generally point in different directions.  

In 3 + 1 space the intersection of a backward and forward time cone will produce a spherical surface like the one pictured in Fig.\ 4c.  In this case choose three mutually perpendicular right-hand unit vectors $\hat{x}$, $\hat{y}$, $\hat{z}$ to form the spatial local grid at event \textbf{a}.  As before, the orientation of these spatial grids is of no importance.  They may be arbitrarily directed because their only purpose is to locally define all three spatial derivatives of the function $\psi$.  That function is continuous throughout the wave packet in any direction; therefore, \emph{it does not matter which grid orientation is chosen at any event} for the purpose of specifying the function and its continuous derivatives at that event.  The Dirac solution has four components $\psi_\mu(\textbf{a})$ where each satisfies all of the above conditions in the 3 + 1 directions.

Since every event on the surface of the sphere in Fig.\ 4c locates a partition line, the event \textbf{a} is enclosed by a sphere with a differential volume $d\Omega$ that contains a differential fraction $df$ of the entire particle, where  
\begin{displaymath}
\psi^*(\textbf{a})\psi(\textbf{a})d\Omega(\textbf{a})=df(\textbf{a})
\end{displaymath}
which normalizes the 3 + 1 wave function.

\section*{Applying the dynamic principle (3 + 1)}
The third condition on a wave function $\psi^(\textbf{a})$ in Eq.\ 1 requires that the dynamic principle applies throughout the space.  This can be done in the 3 + 1 space of an event a by using the grid defined in Fig.\ 4c.  Since we can do this at any event and for any orientation of the grid, we state the more general form of the third condition:

\begin{quote}
\emph{A particle's continuous wave function $\psi(\textbf{a})$ and its derivatives at an event \textbf{a} is subject to a dynamic principle that is applied locally to any four mutually perpendicular space-time directions centered at \textbf{a}, where time is directed along the partition line through \textbf{a}.  This principle determines how $\psi(\textbf{a})$ evolves relative to its own time against the metric background, and how it relates spatially and temporally to its immediate neighbors.  The orientation of local grids are not systematically related to each other; nonetheless, the derivatives of $\psi(\textbf{a})$ defined by these grids will be continuous throughout the particle}.
\end{quote}

\section*{Two particles}
Figure 5 shows the partition lines of two separated massive particles where each has its own definition of a grid that is different from the other particle.  It is a consequence of the trans-coordinate picture that these particles in isolation will seem to have nothing to do with one another.  However, the positional relationship of one to the other is objectively defined in the metric space  of both.  Every event in the wave packet of each particle has a definite location in the metric space, and that fixes the positional relationship of each part of each particle with other parts of itself and with other particles\footnote{This statement is qualified when gravitational uncertainty is considered.  See the second to last section in this chapter.}.   In addition, each particle produces a gravitational disturbance that has an invariant influence on the other particle.  That influence is a function of the relative velocity between the two, even though kinetic energy is not defined for either one.   Kinetic energy is a coordinate-based idea as we have said, whereas the metrical positions and gravitational disturbances in the metric are invariant.  

\begin{figure}[h]
\centering
\includegraphics[scale=0.8]{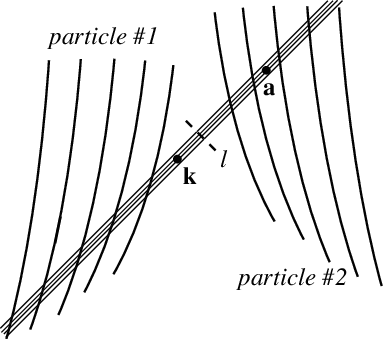}
\center{Figure 5: Two particles and a photon}
\end{figure}

\section*{A radiation photon}
The pack of four lines that rise along the light line in Fig.\ 5 are intended to be the partition lines of a radiation photon with a group velocity $c$.  Photons also have partition lines that separate them into fractional parts, which is a separation by phase differences.  The photon in Fig.\ 5 is confined to the wave packet that is distributed over the perpendicular dashed line $l$.  In empty space its only properties are its phase differences and its spin.  The radiation field is a phase field.

Normally in physics we do not hesitate to use coordinates in empty space, so a photon by itself will be given a period and wavelength relative to that coordinate frame, and hence an energy and momentum.  But if coordinates in empty space have no legitimate place in physics, than like any other particle a photon by itself will lack translational variables (e.g., energy and momentum). It should be clear from the diagram in Fig.\ 5 that the photon bundle has no definable wavelength or frequency at event \textbf{k}, independent of a choice of a world line.  And since the photon has no `internal' energy (i.e., its rest energy is zero) it has no mass contributing to the curvature of space.  The gravitational perturbation of its light line will therefore be zero.  There is no photon mass or energy to perturb it.  These particles move over \emph{light line geodesics} that are created by other (massive) bodies; so although radiation photons are themselves massless and hence weightless, they behave as though they are attracted to gravitational masses.

\section*{Information transfer}
 It is the photon's phases that affect a transfer of energy and momentum from one particle to another.  This is shown in Fig.\ 6 where two particles are defined to be moving over world lines $w_1$ and $w_2$.  The two dashed lines in the figure represent the partition lines of a passing photon with a `relative' phase difference given by $\delta \pi$.  If the photon wave is made up of more than one frequency, the phase difference due to the $i^{th}$ frequency is represented by $\pi_i$. 
 
The photon interacting with the first particle at event \textbf{a} will have a local energy and momentum given by $e_\gamma(\textbf{a}), p_\gamma(\textbf{a})$, and as it interacts with the second particle at event \textbf{b} it will have a local energy and momentum given by $e_\gamma(\textbf{b}), p_\gamma(\textbf{b})$.  These quantities are constructed using the phase relationships that are transmitted between particles, and are articulated in the local grid of the interacting particle. 
\begin{eqnarray}
a \hspace{.1cm} photon \hspace{.1cm} at \hspace{.1cm} event &\textbf{a}:&  e_\gamma(\textbf{a}) = 
\hbar\Sigma_i\omega_i(\textbf{a}) \hspace{.5cm}  p_\gamma(\textbf{a}) = \hbar\Sigma_ik_i(\textbf{a}) \nonumber \\
a \hspace{.1cm} photon \hspace{.1cm} at \hspace{.1cm} event  &\textbf{b}:&  e_\gamma(\textbf{b}) = 
\hbar\Sigma_i\omega_i(\textbf{b}) \hspace{.5cm}  p_\gamma(\textbf{b}) = \hbar\Sigma_ik_i(\textbf{b}) \nonumber
\end{eqnarray}
where $\omega_i(\bf{a})$ = $\partial_t\pi_i(\bf{a})$ and $k_i(\bf{a})$ = $\partial_x\pi_i(\bf{a})$. These derivatives refer to the local grid of each event in each particle and are defined like those in Eq.\ 2.  So a photon at an emission site \textbf{a} or an absorption site \textbf{b} will have a well-defined energy and momentum at those sites, although it does not have these properties during the flight from \textbf{a} to \textbf{b}.  
\begin{figure}[t]
\centering
\includegraphics[scale=0.8]{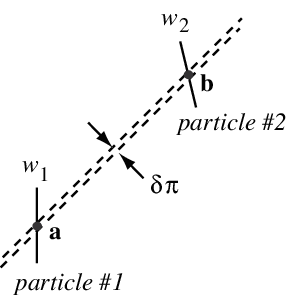}
\center{Figure 6: Two particles and a photon}
\end{figure}

\section*{Relativistic symmetry}
In a Newtonian inertial system the space-time symmetries were described in terms of the behavior of clocks and meter sticks.  That kind of description is not so easy in the relativistic case, so we put the matter differently.  The basic symmetry in special relativistic systems is to be found in the structure in the background invariant metric.  \emph{Every event in an inertial system is located at the origin of a Minkowski metric with the same temporal and spatial intervals throughout the space}, and with future and past time cones bordered by $45^\circ$ \mbox{light lines}.

If the world line of particle \#1 in Fig.\ 6 is parallel with that of particle \#2 in an inertial space with this symmetry, then the energy and momentum at event \textbf{a} will be identical with the energy and momentum at event \textbf{b}, so these quantities will be conserved at these two events.  However, these quantities will still not be defined \emph{between} these two events.  

If the particleÕs world lines were not parallel, then these quantities would not identical at particles \#1 and \#2.  We would then say that there has been a Doppler shift, although we cannot calculate that shift because the relative velocity of the particle is not defined.  Of course nature does not need to calculate the Doppler shift because the shift is fully determined by the way that the first world line is related to the second world line.  Only we feel a need to do a computation like that, and that is one of the reasons we use coordinates -- to assign velocities to particle world lines in order to find the Doppler shift in terms of the relative velocity between them.  Another reason is that coordinates let us say that the photon's energy and momentum are the same for the entire trip from \textbf{a} to \textbf{b}.  Establishing a vertical time line and a perpendicular horizontal direction at every event along the way in Fig.\ 6 makes it possible to specify the period and wavelength of the photon (hence its energy and momentum) during its flight time from \textbf{a} to \textbf{b}.  The addition of coordinates therefore facilitates analysis and allows us to extend the conservation principles throughout the space, thereby giving us \emph{continuous conservation}.

A special relativistic system without coordinates evidently works very much like a Newtonian system without coordinates.  Energy, momentum, and angular momentum are conserved in local interactions, and they are conserved for two or more separated interactions because the background metric has the right symmetries.  However, nature in trans-coordinate physics does not provide for continuous conservation.  It does not conserve energy, momentum, and angular momentum at every moment of time.  Nature has no need to do that.  We are the ones who do that.  We add coordinates because we have a need to calculate things (such as Doppler shift) in terms of other things, and to conserve as much as possible at every moment of time.  Adding coordinates aids analysis and adds immeasurable to the big picture but they are a construction of our own making.

Symmetries in a more generalized geometry are not plentiful.  There are always the Minkowski symmetries in the tangent plane at every event, and there are special situations like the spherical symmetry of a Schwarzschild metric.  But there is no general symmetry that characterizes the generalized metric because it \emph{is} a generalized metric.  When coordinates are added it is possible to identify symmetries by taking derivatives of the metric tensor.  If $g_{\mu\nu}$ is time independent then energy is conserved.  If that tensor is independent of any spatial coordinate then momentum in that direction is conserved, and if it is independent of rotation then angular momentum is conserved in that \mbox{direction \cite{RM1}}.  In a general geometry there is no guarantee that one can always find coordinates that conserve these quantities without introducing pseudo-tensors for that \mbox{purpose \cite{LL}}.

\section*{Classical electromagnetic radiation}
In classical three dimensional coordinate-based physics the  \emph{electromagnetic potential}  of a radiation field is given by a fourvector $A^\mu(\textbf{a})$, where the d'Alembertian operating on $A^\mu(\textbf{a})$ is equal to zero.  However, trans-coordinate physics cannot use the d'Alembertian in empty space.  Where a particle grid exists we can give analytic expression to that dynamic principle; but where there is no grid we can only specify the phases of a `polarized' $A$-field.  These phases will appear in \mbox{(3 + 1)} Minkowski space in parallel relationships that are similar to those in the (1 + 1) diagrams in Figs.\ 5 and 6.  However, the \emph{vector nature} of the wave does not appear until it interacts with matter that provides the necessary grid.  The dynamic principle can then be applied to the resulting $A^\mu$ components, with derivatives defined in a four-dimensional version of Eq.\ 2.

 Any photon-based radiation must be qualified in a way that provides for Huygens' wavelets.  So far we have assumed that this radiation moves in an outward direction from one original source along a light cone.  We now say that any event along the way (such as \textbf{k} in Fig.\ 5) will act as a source of radiation in all directions.  The wavelet from an event \textbf{a} has the same phase and polarization as the primary wave at event \textbf{a}, and radiates uniformly in all directions with a velocity $c$.  Two wavelets that arrive at another event \textbf{b} will have a definite phase difference that produces interference there.

\section*{Virtual photons}
In standard coordinate physics an isolated stationary charge is supposedly surrounded by a scalar Coulomb field $\phi$.  In a moving frame the charge is also said to have a vector field \textbf{A} in the direction of motion.  In trans-coordinate physics we say that there is no such field.  There is only a single charge that moves along a given world line.  Virtual photons manifest themselves only when there are at least two interacting charges.  

Assume that we have two adjacent charged particles that are classically well localized.  There are two space-like interaction events in this case, one involving a virtual photon and the first particle, and another involving a virtual photon and the second particle.  These interactions are in the nature of correlated non-local events with nothing in the space-like interval between them but photonic phase relationships like a radiation photon.    And similarly, a virtual photon carries no grid of its own, so each of these interactions employs the grid of the interacting charged particle to define each energy-momentum transfer.

Impose a \emph{common inertial frame} on two stationary charged particles.  The time $t$ assigned to event grids in the first particle and the time $t'$ assigned to event grids in the second particle can be set equal to one another other and to the time of the common inertial frame.  Also when using these coordinates, it does no harm to assume that the spatial grid at each event of each particle in the system is aligned with the spatial variables of the common coordinate frame.  The virtual exchange at each end of this interaction will then produce \mbox{a Coulomb} intensity that is equal to $(q_1q_2/4\pi r)\delta(t - t')$, where $r$ is the \mbox{distance between} the particles in the common frame, and $t$ equals $t'$ in the common frame.  In another Lorentz frame  \emph{relativistic corrections to this equation occur when the spatial components of the current four-vectors are taken into account} according to Feynman \cite{FMW}.  So from a coordinate point of view, each of these charges (in motion or not) will produce the conventional $\phi$ and \textbf{A} fields at the location of the one another.  We say that these potentials appear on the grids of each of these particles with only polarized phase relationships appearing in the space between them.

We generalize this result by saying that each charged particle in a group of charged particles (in motion or not) will produce the classical values of $\phi$ and \textbf{A} at each of the other charged particles.  Using the local grid at an event \textbf{a} inside the wave function of one of these particles, it is possible to take local derivatives (using a four-dimensional Eq.\ 2) to find the classical fields \textbf{E} and \textbf{B} as well.  None of these vector fields exist in the empty space between the particles.

Example:  The electromagnetic field \textbf{B} in the empty space around an electrically neutral current carrying solenoid is equal to zero.  However, the electric charges that make up the current are a  source of virtual photons that can influence particles that have been placed inside the solenoid; so when a neutron is placed at an event \textbf{a} inside the solenoid its grid allows the vector field \textbf{A} to be defined at \textbf{a} that is tangent to a circular path centered on a cross section of the solenoid.  Although grids may be differently oriented for events in the immediate neighborhood of \textbf{a} inside the neutron, this does not affect the continuousness of the field \textbf{A} or its derivatives in the neutron.  Therefore, the curl of \textbf{A} and the magnetic field \textbf{B} are well defined at event \textbf{a}, and both are continuous from one  neighborhood to the next.  And because of the continuousness of the neutron's wave function, its spin is aligned with \textbf{B} in the same way from one neighborhood to the next.  

With virtual photons we introduce another regional influence.  We first introduced the invariant metric background that is the underlying canvas for everything we do.  This defines the region that is occupied by the system.  We then introduced partition lines that cover the region occupied by a particle and allowed us to define a continuous function $\psi$ over its volume.  And now we introduce virtual photon interactions that are non-local influences between particles.  However, we have always applied the dynamic principle `locally' and that continues to be the case during an interaction.

When two charged particles interact, every event of each particle is affected by every event of the other particle.  The result is a particle-wide shift in the partition lines of each particle, changing each particle's function $\psi$ from its pre-interaction value to something new.  The Hamiltonian including the interaction term is now applied \emph{locally} to every event in both particles, establishing a site of virtual exchange, and directing the local response to this \emph{non-local} interaction.

\section*{General Relativity}
When the grid of an event \textbf{a} inside a particle wave function is combined with the metric background, it is possible to define a metric tensor $g_{\mu\nu}$ at \textbf{a}.  However there are an infinite number of ways that a continuous $g_{\mu\nu}$ field `might' be defined in the neighborhood around \textbf{a}, corresponding to the infinite number of coordinate systems that `might' be chosen in that neighborhood. Therefore, physical significance cannot be attached to a continuous metric tensor in a trans-coordinate universe.  An exception is the component $g_{44}(\textbf{a})$.  This expresses the relationship between the unit vector $\hat{t}$ at event \textbf{a} and the invariant interval in the time direction, and since that relationship is continuous throughout the particle's wave function, the component $g_{44}$ has time derivatives defined at each event, using Eq.\ 2.  However, \emph{none} of the derivatives of $g_{\mu\nu}$ are defined in empty space, so Christoffel symbols are not generally defined.  Parallel displacement is generally meaningless. An affine connection does not exist.

It follows that the Riemann and Ricci tensors are not generally defined; and as a result, the field equation of general relativity is not generally defined.  However, the gravitational field equation can generate coordinate solutions that fit Riemannian surfaces in special situations, and that makes it very useful.   For instance, the metric space around a stationary spherical mass possesses temporal and spherical symmetry, and that implies the existence of conserved quantities in that space.  The Schwarzschild coordinate solution fits that surface in the same way that Cartesian coordinates fit a Newtonian surface.  The Schwarzschild coordinates therefore provide a way of analyzing object behavior and giving numerical expression to continuously  conserved quantities.  In particular it is possible to identify the geodesics of satellites whose masses are sufficiently small as to not perturb the metric in a significant way, and to provide for the  conservation of energy and angular momentum for those objects over the length of their geodesics.  

Like a special relativistic system, a general relativistic system without coordinates works very much like a Newtonian system without coordinates. Conservation principles apply where there is sufficient symmetry, but they do not apply continuously over geodesics between interactions.  Nor can we attribute velocity to objects on geodesic paths, with the exception of photons that will always move with the `local' velocity of light.  Schwarzschild coordinates do what other coordinate systems do.  They let us find conserved quantities over the length of an object's geodesic and allow us to give numerical expression to those quantities.    They give us the big picture and let us make analytic connections in a way that would not otherwise be possible \mbox{-- connections} that nature has no need make.

If general relativity is not the fundamental gravitational theory, then we must turn to graviton theory.   Gravity  must then be represented in some way by quantum mechanics.  We say that classical general relativity is a science that only approximates an underlying quantum reality.

\section*{Gravitational uncertainty}
Throughout this chapter it has been assumed that the invariant metric space has definite values in the classical sense.  But if  gravitons are the exchange particles of ordinary gravity, then their influence on metric space will have associated uncertainties.  The question is: How does everything we have said so far about particle dynamics survive these uncertainties?

We will assume that the metric relationship between an event \textbf{a} to another event \textbf{b} is generally uncertain. However, any small neighborhood of event \textbf{a} will retain the structure shown in Fig.\ 2.  That is, both events \textbf{a} and \textbf{b} in that figure will generally be uncertain, but the relationship between them will take on a definite value in the limit as -$\Delta$ goes to zero.  Because the velocity of light is equal to $c$ in that limit, events \textbf{c} and $\textbf{c}'$ in that figure will also have a definite metrical relationship to each other.  So for small volumes, metric space retains a definite Minkowskian structure.  This is similar to the case of an atom whose wave function has grown to a large volume (because of its uncertainty of momentum) relative to its `initial' volume.  Even with this expanded uncertainty of position, the atom's structure will be preserved at each location; it will retain its characteristic energy eigenstates independent of its uncertainty of position.  In the same way, a metric space that is wildly uncertain over finite distances will nonetheless have a definite Minkowski structure in the infinitesimal region around any particular event.

A particle's wave function $\psi(\textbf{a})$ at event \textbf{a} will also be uncertain, but any particular value $\psi'(\textbf{a})$ will be correlated with other values of $\psi'$ located at other events throughout the particle.  These correlations occur because the uncertainty in the metric leads to an uncertainty in the spacing of the partition lines, and this leads to an uncertainty in the square modulus associated with any event in the wave function of the particle.  Therefore, a particular value $\psi'(\textbf{a})$ does not exist by itself.  It involves all the other correlated values of $\psi'(\textbf{a})$ that are related to each other by a \emph{specific spacing of the partition lines}.  We therefore identify a particular regional function $\psi'$ that extends over the particle and is continuous with $\psi'(\textbf{a})$ as specified in Eq.\ 1.  

It follows from its continuousness that each value of $\psi'(\textbf{a})$ will have definite derivatives defined for it by the limiting process of trans-coordinate physics; so  when the dynamic principle is applied to that event, each value of $\psi'(\textbf{a})$ will evolve as though it were the only value of $\psi$ at that event.  In trans-coordinate physics, local dynamic evolution is all that is defined or is required of an isolated particle.

When two particles interact a non-local correlation is established that does not really change this picture for reasons that were previously stated.  The interaction causes a particle-wide shift in the partition lines of each particle and thereby changes any particular function $\psi'$ of one of the particles to something new, call it a particular continuous function $\psi''$.  As in the previous section on virtual photons, the Hamiltonian including the interaction term is applied `locally' to every event in both particles.  This establishes each event as a site of virtual exchange, and directs the local response of each $\psi''$ as though it were the only $\psi''$ at that event.
   
I believe that these conditions are necessary to reconcile gravitational uncert-ainty with the dynamics of particles that follow well-defined differential equations.  This relationship is a severe problem in coordinate physics because the goal there is to join these ideas over finite regions of space, including the artificial framework of the coordinates themselves.  However, if the dynamic principle is \emph{entirely locally contained} without an unnatural coordinate superstructure as it is in trans-coordinate physics, then every local event in every finite region of space will take care of itself.  It will be dynamically well behaved in its own neighborhood and in its relationship to its immediate neighbors.  The whole will be the sum of these parts, independent of the gravitational curvature and uncertainty thereof.

\section*{Gravity waves}
	Gravity waves are said to cause the space occupied by material things to expand and contract, so the diameter of a neutron will expand and contract along perpendicular axes as a gravitational wave passes over it.  These waves are represented in general relativity as variations in the metric tensor $g_{\mu\nu}$ that propagate through the empty space between the source and the detector.  However, trans-coordinate physics does not allow a metric tensor to be defined in empty space.  Therefore, gravity waves as understood by general relativity do not exist.  

	This limitation did not deter us from using photonic phases and polarizations to cross the gap between a source and a receiver, thereby delivering the electromagnetic goods.  Maybe gravitons can do the same for gravity, delivering the expected effects to gravity wave detectors like the Weber bar and the LIGO interferometer.  The trouble is that the cause and the effect are not compatible with one another in the gravitational case as they are in the electromagnetic case.

	The presumed cause of gravity waves is a pair of rotating masses in a circular or an elliptical orbit, or it is a pair of masses that blow apart or fall together.  The question is: What part of a circular or elliptical path corresponds to the expansion of an affected neutron many miles away, and what part corresponds to its contraction?  Or, what part of two masses falling together or blowing apart corresponds to the expansion of the affected neutron, and what part corresponds to its contraction? A rotating or accelerating cause simply does not go together with the pulsating effect predicted by general relativity.  This is not like the electromagnetic case in which the activity at the receiver parallels the (retarded) polarization and activity at the source.  Therefore, a direct transfer of energy/momentum is not possible because the claimed gravitational `cause' and the supposed `effect' are a complete mismatch.

It follows energy and momentum are carried away from a radiating source by granular gravitons rather than continuous waves.  Neither a Weber bar nor a LIGO interferometer will respond to this radiation because these instruments were not designed to detect gravitons. 

\begin{center}
\line(1, 0){130}
\end{center}

In  is shown in Ref.\ 6 how a trans-coordinate system of particles in \mbox{3 + 1} space can be represented as a state, and how a Hamiltonian can be written for that state.  It also describes the invariant nature of a quantum mechanical ÔcollapseÕ of one of these states, which is the fourth regional influence.


\begin{thebibliography}{99}

\bibitem{tH}'t Hooft, G.    ``A Grand View of Physics", \emph{Int'l J. Mod. Phys.} \textbf{A23} sect 3,  p. 3755 (2008); arXiv:0707.4572 


\bibitem{JM}Maldacena, J.   "The Illusion of Gravity", \emph{Sci. Am.} Nov, p. 56 (2005)


\bibitem{RM1}R. A.Mould, ``Basic Physics", Springer, New York (2002) Eq. 8.66



\bibitem{LL}Landou, L. and Lifshitz, E.  \emph{The Classical Theory of Fields}, Pergamon Press, New York, (1971) p. 316

\bibitem{FMW}  Feynman, R. P.,  Morinigo, F. B., Wagner, W. G.  \emph{Feynman Lectures on Gravitation},    B. Hatfield. ed., Addison-Westley, New York, 33 (1995)


\bibitem{RM2}R. A.Mould, ``Trans-Coordinate States", arXiv:0812.1937


	






\end{thebibliography}
\end{document}